# Analytic Interatomic Forces in the Random Phase Approximation

Benjamin Ramberger, Tobias Schäfer, and Georg Kresse[*]

University of Vienna, Faculty of Physics and Center for Computational Materials Sciences,
Sensengasse 8/12, 1090 Vienna, Austria

We discuss that in the random phase approximation (RPA) the first derivative of the energy with respect to the Green's function is the self-energy in the $GW$ approximation. This relationship allows us to derive compact equations for the RPA interatomic forces. We also show that position dependent overlap operators are elegantly incorporated in the present framework. The RPA force equations have been implemented in the projector augmented wave formalism, and we present illustrative applications, including *ab initio* molecular dynamics simulations, the calculation of phonon dispersion relations for diamond and graphite, as well as structural relaxations for water on boron nitride. The present derivation establishes a concise framework for forces within perturbative approaches and is also applicable to more involved approximations for the correlation energy.

Density functional theory (DFT) is almost ubiquitous in present day computational chemistry and materials science. Compared to other methods, one of the great advantages of DFT is that forces between the atoms are readily computable, so that relaxed ground state structures, vibrational properties, as well as finite temperature properties of any material are straightforwardly accessible. Although quantum chemists have devised ways to compute analytic forces, e.g., for Møller-Plesset perturbation theory (MP) already decades ago [1,2], in condensed matter simulations forces beyond DFT have been rarely available. This has limited the success of more involved electronic structure methods such as the random phase approximation (RPA) for solids. Only recently, this situation has improved, with first derivatives being computable for second order MP perturbation theory (MP2) for solids [3], as well as for the RPA for molecules [4,5]. However, to date all these implementations are limited to real valued Gaussian-type orbitals and even then the algebra is often intricate.

Here we use Green's function theory, to derive compact equations for the forces in the random phase approximation. These can be recapped in two lines. One first calculates a Hermitian first order one-particle density matrix

$$\rho^{(1)} = \frac{1}{2\pi}\int_{-\infty}^{\infty} d\nu G(i\nu)[F + \Sigma_{GW}(i\nu)]G(i\nu), \quad (1)$$

involving the Fock operator $F$, the correlation part of the self-energy in the $GW$ approximation $\Sigma_{GW}$, and the independent particle Green's function $G(i\nu)$ of the Kohn-Sham (KS) Hamiltonian. Then one needs to trace this density matrix over the first derivate of the KS Hamiltonian $F_i = -\text{Tr}[\rho^{(1)}\partial H_{KS}/\partial R_i]$, where $R_i$ are the ionic coordinates and $F_i$ is the corresponding component of the force. We have implemented the corresponding equations in the projector augmented wave [6,7] formalism and show the potential of this method in the present Letter.

The fundamental problem we address here is that perturbation theory (including the RPA) is nonvariational; i.e., the kinetic, Hartree, exchange, and correlation energies are evaluated using a set of orbitals calculated using some other zero-order Hamiltonian. For the purpose of this presentation, we assume that these one-electron orbitals are always determined using a KS Hamiltonian $H$. We rely on Green's function theory and establish the essentials first and then proceed to an outline of the derivation. Given the generalized eigenvalue equation

$$H|n\rangle = \epsilon_n S|n\rangle, \quad (2)$$

with a KS Hamiltonian $H$ and an overlap operator $S$, the time dependent Schrödinger equation reads

$$(iS\partial/\partial t - H)|n\rangle = 0. \quad (3)$$

Then the single particle KS Green's function $G$ can be written as

$$G(i\nu) \coloneqq (i\nu S - H)^{-1}. \quad (4)$$

In the present Letter, the Green's functions are represented in imaginary frequency $i\nu$ and time $i\tau$, where they are smooth functions of their arguments. The Fermi-level is placed at $\mu = 0$; occupied (unoccupied) states have all negative (positive) one-electron energies.

Green's function theory allows us to write compactly the change of the Green's function upon perturbations. This is achieved by the Dyson equation, which relates the perturbed Green's function $G + \Delta G$ corresponding to $H + \Delta H$ and $S + \Delta S$ to the unperturbed Green's function $G$:

$$(G + \Delta G)^{-1} = G^{-1} - (\Delta H - i\nu \Delta S) \Leftrightarrow$$
$$(G + \Delta G) = G + G(\Delta H - i\nu \Delta S)(G + \Delta G). \quad (5)$$

Neglecting $\Delta G$ on the rhs of the second line, one obtains for the change of the Green's function $\Delta G$ the first order term

$$G^{(1)}(i\nu) = G(i\nu)(\Delta H - i\nu \Delta S)G(i\nu). \quad (6)$$

It is noted that $\Delta H$ is the total change of the KS Hamiltonian including the self-consistent field changes.

The Green's function in imaginary time can be obtained by a Fourier transformation:

$$G(i\tau) = \frac{1}{2\pi} \int_{-\infty}^{\infty} G(i\nu) e^{-i\nu\tau} d\nu. \quad (7)$$

Using standard mathematical relations, it is easy to show that one can also write [8,9]

$$G(i\tau) = \begin{cases} \sum_{j \in \text{occ}} |j\rangle\langle j| e^{-\tau \epsilon_j} & \tau < 0 \\ -\sum_{a \in \text{virt}} |a\rangle\langle a| e^{-\tau \epsilon_a} & \tau > 0. \end{cases} \quad (8)$$

Here the sums are restricted to the occupied (occ) manifold for negative $\tau$ and to the unoccupied or virtual (virt) manifold for positive $\tau$, and thus the exponents are always bound. Clearly [see Eq. (8)], the one-particle density operator $\rho$ is given by the Green's function for $\tau \to 0^-$, i.e., for a slightly negative imaginary time

$$\rho = \lim_{\tau \to 0^-} G(i\tau) = G(i\eta), \quad (9)$$

where $\eta = 0^-$ is a small negative infinitesimal.

We now proceed to calculate first derivatives of the Hartree-Fock energy functional. This functional depends on the density matrix $\rho$ and an external potential $V_{\text{ext}}$: $E = E_{\text{HF}}[\rho, V_{\text{ext}}]$. The total variation is

$$\delta E_{\text{HF}} = \underbrace{\text{Tr}\left(\frac{\delta E_{\text{HF}}}{\delta V_{\text{ext}}} \delta V_{\text{ext}}\right)}_{\text{Hellmann-Feynman}} + \underbrace{\text{Tr}\left(\frac{\delta E_{\text{HF}}}{\delta \rho} \delta \rho\right)}_{\text{non-Hellmann-Feynman}}. \quad (10)$$

The trace operator usually involves integration over one (for the variations of the external potential) or two spatial coordinates (for variations of the density matrix). If one calculates the HF energy derivatives at the HF density matrix $\rho_{\text{HF}}$, terms related to variations of the density matrix do not need to be calculated, since the Hartree-Fock density matrix fulfills the equation

$$\text{Tr}\left[\frac{\delta E_{\text{HF}}}{\delta \rho} \delta \rho\right]_{\rho = \rho_{\text{HF}}} = \text{Tr}[F \delta \rho]_{\rho = \rho_{\text{HF}}} = 0, \quad (11)$$

where $\delta\rho$ is subject to the orthonormality constraint of the corresponding basis functions and $\delta E_{\text{HF}}/\delta \rho$ is the Fock operator $F$. In this case the second term in Eq. (10) vanishes, which is just the famous Hellmann-Feynman theorem.

If the random phase approximation is used, the Hartree-Fock energy functional, however, needs to be evaluated at the KS density matrix. The corresponding energy is commonly referred to as exact exchange energy (EXX) [10]. The non-Hellmann-Feynman term is then nonzero. Using Green's functions, we can write this term compactly by combining Eqs. (6), (7), and (9). For the non-Hellmann-Feynman term in Eq. (10) one then obtains

$$\frac{1}{2\pi} \int_{-\infty}^{\infty} d\nu \text{Tr}[FG(i\nu)(\delta H - i\nu \delta S)G(i\nu) e^{-i\nu\eta}], \quad (12)$$

where the Fock operator $F$ is constructed from the KS density matrix. Equation (12) looks expensive to evaluate, since, seemingly, for any change of the Hamiltonian $G(i\nu)\delta H G(i\nu)$ needs to be calculated. However, since the trace is invariant under cyclic permutations, one can calculate two Hermitian matrices,

$$\rho_{\text{HF}}^{(1)} = \frac{1}{2\pi} \int_{-\infty}^{\infty} d\nu G(i\nu) F G(i\nu) e^{-i\nu\eta},$$
$$\gamma_{\text{HF}}^{(1)} = \frac{i}{2\pi} \int_{-\infty}^{\infty} d\nu \nu G(i\nu) F G(i\nu) e^{-i\nu\eta}, \quad (13)$$

once and then efficiently calculate the energy change for any variation of the KS Hamiltonian and overlap operator as

$$\frac{\partial E}{\partial R_i} = \text{Tr}\left[\rho^{(1)} \frac{\partial H}{\partial R_i} - \gamma^{(1)} \frac{\partial S}{\partial R_i}\right]. \quad (14)$$

We note that $\rho^{(1)}$ is a density matrix, whereas $\gamma^{(1)}$ has the dimension of an energy (first moment $i\nu$). Furthermore, since $F$ is frequency independent, the integrals in Eq. (13) can be calculated analytically using the residue theorem, where the term involving $\eta = 0^-$ specifies how the contours need to be closed (see Supplemental Material [11]).

This equation can be straightforwardly generalized to correlation energy functionals that depend on the Green's function $E_c = E_c[G]$. Since the correlation energy usually does not depend on the external potential, the Hellmann-Feynman term is zero for this contribution. The non-Hellmann-Feynman term yields an additional contribution to the matrices,

$$\rho_c^{(1)} = \int_{-\infty}^{\infty} d\nu G(i\nu) \Sigma_c(i\nu) G(i\nu),$$
$$\gamma_c^{(1)} = i \int_{-\infty}^{\infty} d\nu \nu G(i\nu) \Sigma_c(i\nu) G(i\nu), \quad (15)$$

where the self-energy $\Sigma_c$ is the functional derivative of the correlation energy with respect to the Green's function $\Sigma_c(i\nu) = \delta E_c/\delta G(i\nu)$. Equation (15) is obviously analogous to Eq. (13), only the Fock operator $F$ has been replaced by the self-energy. Furthermore, the exponential $e^{-i\nu\eta}$ can be dropped, since the integrand decays like $o(1/\nu^2)$, and the integral is unconditionally convergent.

For the correlation energy in the random phase approximation, $E_{\mathrm{RPA}}$, the corresponding self-energy is exactly the correlation part of the $GW$ approximation to the self-energy in Hedin's approach [12], which we rationalize now. The RPA correlation energy is defined as [13–19]

$$E_{\mathrm{RPA}} = \frac{1}{2\pi}\int_0^\infty d\nu \mathrm{Tr}[\ln(1-\chi(i\nu)v) + \chi(i\nu)v], \quad (16)$$

where $\chi(i\nu)$ is the independent particle polarizability calculated using KS orbitals, and $v$ is the Coulomb kernel. The derivative with respect to $\chi(i\nu)$ can be calculated by Taylor expanding the logarithm and taking the derivative (see also Ref. [20]). This yields the correlation part of the screened Coulomb potential

$$-\frac{\delta E_{\mathrm{RPA}}}{\delta \chi(i\nu)} = W_c(i\nu) = v\chi(i\nu)v + v\chi(i\nu)v\chi(i\nu)v + \ldots$$

Using the relation between the independent particle polarizability and the Green's function [12]

$$\chi(\mathbf{r},\mathbf{r}',i\tau) = G(\mathbf{r},\mathbf{r}',i\tau)G(\mathbf{r}',\mathbf{r},-i\tau), \quad (17)$$

one formally obtains the desired relation (dropping the position coordinates):

$$\frac{\delta E_{\mathrm{RPA}}}{\delta G(i\tau)} = \frac{\delta E_{\mathrm{RPA}}}{\delta \chi(i\tau)}\frac{\delta \chi(i\tau)}{\delta G(i\tau)} = -W_c(i\tau)G(-i\tau)$$
$$= \Sigma_{GW}(i\tau). \quad (18)$$

Details are given in the Supplemental Material [11].

In summary, non-Hellmann-Feynman forces are simple to evaluate. One only needs to determine the one particle density matrix $\rho^{(1)}$ and its first energy moment $\gamma^{(1)}$, Eqs. (13) and (15), and then evaluate Eq. (14). The first derivative of the self-consistent density functional theory Hamiltonian $\partial H/\partial R_i$ and of the overlap operator $\partial S/\partial R_i$ need to be calculated for each ionic displacement, $\delta R_i$, yielding one component in the atomic forces.

Concerning the scaling, we note that we have recently implemented an algorithm that calculates the $GW$ self-energy with linear scaling in the number of $k$ points and cubic scaling in the number of plane waves [21] following the work of Rojas, Godby, and Needs [22,23]. The algorithm is designed to use a minimal number of time and frequency points, with 12 points yielding $\mu$eV accuracy per atom for gapped systems [24,25]. The scaling with the number of plane waves and $k$ points carries over to the calculation of the density matrices, so that the complexity of the entire calculation is, in principle, identical to standard DFT, but with much larger prefactors. In practice, one force calculation requires about 3 orders of magnitude more compute time than for a semilocal functional, about 10–50 times longer than for hybrid functionals (albeit scaling quadratically in the number of $k$ points), and about 10 times longer than for a single RPA total energy calculation. One reason for the steep computational cost is that the calculation of the $GW$ self-energy requires about 6 times more operations than an RPA total energy calculation. Furthermore, the evaluation of $\partial H/\partial R_i$ for each ionic displacement is currently done using linear response theory and can take about the same time as the calculation of $\rho^{(1)}$ and $\gamma^{(1)}$ for systems with hundreds of atoms. On the positive side, $G_0W_0$ quasiparticle energies for all orbitals are readily available and can be calculated *en passant*.

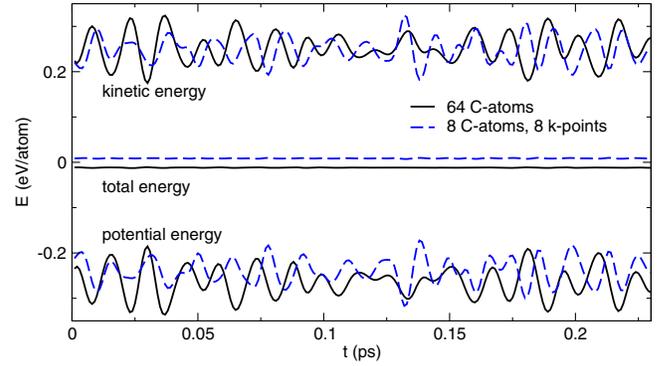

FIG. 1. Potential and kinetic energy per atom for diamond supercells of 64 and 8 atoms, using the $\Gamma$ point and $2\times 2\times 2$ $k$ points, respectively. Energies are shifted so that the total energies approach zero. The slight fluctuations in the total energy are similar to those observed in DFT for a 1 fs time step using a Verlet algorithm.

We have implemented the RPA forces in the projector augmented wave (PAW) [6,7] based Vienna *ab initio* simulation package (VASP). Typically we can obtain 5 digits accuracy compared to a numerical differentiation, and the precision increases as the number of time points increases (see Supplemental Material [11]). A rigorous test for the validity of forces are first principles molecular dynamics (MD) simulations, since errors in the forces will result in a drift of the total energy. In Fig. 1 we show results for short simulations of crystalline diamond at 2000 K. The configurations were initially equilibrated using DFT, and then about 250 RPA-MD steps were performed. For the 64 atom supercell, one time step took 20 min on 40 Intel Xeon v3 cores. It is clear that the energy conservation is excellent for both the small 8 atom diamond cell, as well as the 64 atom cell.

Relaxations to the ionic ground state structure are also feasible and fairly efficient using the present code. One advantage of the present approach is that the second ionic derivatives of the density functional $E_{\mathrm{KS}}$, that is used to calculate the orbitals, are calculated when the linear response code is used. This implies that the KS Hessian matrix $A$ is available,

$$A_{ij} = \frac{\partial^2 E_{KS}}{\partial R_i \partial R_j},$$

and this Hessian can be used as the preconditioner for the update of the ionic positions $\mathbf{R}$

$$\mathbf{R} \leftarrow \mathbf{R} - A^{-1}\mathbf{F},$$

where $\mathbf{F}$ are the RPA forces. If the RPA Hessian is close to the KS Hessian, convergence to the ionic RPA ground state is quadratic and hence, in practice, the forces often decrease by an order of magnitude in each step.

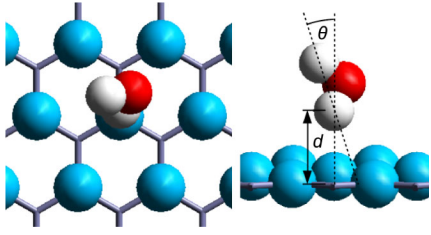

FIG. 2. RPA relaxed geometry for $H_2O$ adsorbed on BN. A $3 \times 3$ unit cell and $3 \times 3 \times 1$ $k$ points were used.

Being able to relax structures adds a new important facet to the modeling of materials beyond density functional theory. Here we have looked at one prototypical problem, $H_2O$ adsorption on a BN layer [26,27]. We have relaxed the $H_2O$ molecule using the RPA and find after few steps a low symmetry configuration atop the N atom that was previously not considered. The main reason for finding this configuration was the use of the DFT Hessian as preconditioner which results in a rapid rotation of the molecule along a very soft mode. All used functionals confirmed that this position is lower in energy than the previously considered structures (6 meV in the RPA). In Table I, we summarize the geometry and adsorption energy. The equilibrium distance to the surface decreases with increasing adsorption energy, with PBE and HSE yielding too small adsorption energies and too large distances. The van der Waals (vdW) corrected functionals generally yield larger adsorption energies and shorter distances than RPA. We expect the RPA to be accurate here, since RPA parameterized force fields yield excellent contact angles for water droplets [27]. However, RPA generally overestimates the Pauli repulsion, with singles increasing the adsorption energy by 20 meV [28].

To demonstrate that the random phase approximation yields also very good vibrational properties, we have calculated the phonon dispersion relation of diamond and graphite (see Fig. 3). We note that the random phase approximation yields excellent structural properties for both, diamond and graphite, with lattice constants in virtually perfect agreement with experiment [35,36]. For the phonons, agreement between the RPA and experiment is also excellent for diamond and very good for graphite. Although phonon dispersions from DFT (not shown) can be in very good agreement with experiment, they are often under- (semilocal functionals) or overestimated (hybrid functionals) [37,38].

In summary, we have presented an elegant scheme to calculate interatomic forces using many body perturbation theory. The derivation relies on Green's functions and the self-energy and seamlessly incorporates position dependent overlap operators. Although applied to the RPA here, analogous expressions can easily be derived for, e.g., MP2, with the RPA self-energy $\Sigma_{GW}$ simply replaced by the self-energy in second order $\Sigma^{(2)}$. Our derivation involves a first order one-particle density matrix, and we note that seemingly related density matrices can be found in the quantum chemistry literature but without reference to a self-energy [1,2,4,5,44]. We have put our equations to the test by implementing them in a plane wave code, and we have demonstrated that MD simulations, relaxations, and predictions of phonon frequencies are possible. We are convinced that the present work paves the way towards accurate beyond DFT modeling in condensed matter physics and materials sciences.

TABLE I. Considered functionals with hydrogen to surface distance $d$ and angle $\theta$ between the H-H axis and the surface normal (see Fig. 2) and adsorption energy $E$.

|  | $d$(Å) | $\theta$(°) | $E$(meV) |
| --- | --- | --- | --- |
| PBE (Perdew-Burke-Ernzerhof) [29] | 2.48 | 18 | 40 |
| HSE (Heyd et al.) hybrid functional [30] | 2.49 | 17 | 34 |
| D3 (Grimme dispersion correct.) [31] | 2.34 | 19 | 142 |
| TS (Tkatchenko-Scheffler) [32] | 2.33 | 19 | 152 |
| MBD (many body dispersion) [33] | 2.33 | 18 | 136 |
| optPBE, vdW DFT functional [34] | 2.31 | 14 | 132 |
| RPA (random phase approximation) | 2.39 | 16 | 92 |

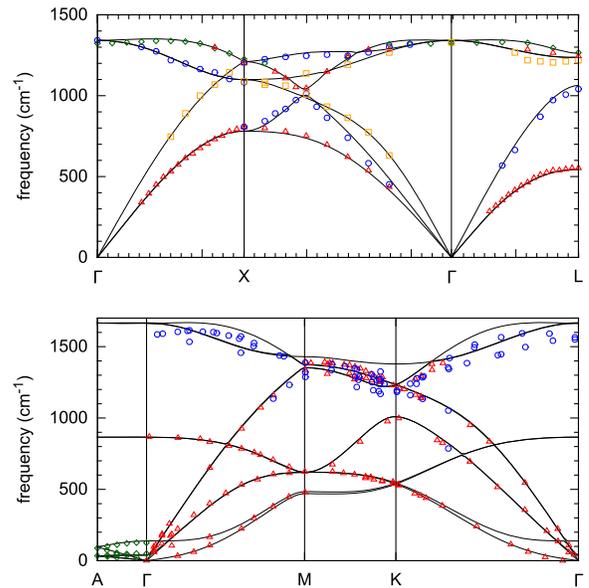

FIG. 3. Phonon dispersion relation of diamond and graphite calculated using 128 atom supercells. RPA calculations were performed at the $\Gamma$ point, with $k$-point sampling errors corrected for using DFT. Experimental data and lattice constants are taken from Refs. [39–43].

Funding by the Austrian Science Fund (FWF): F41 (SFB ViCoM) is gratefully acknowledged. Computations were performed on the Vienna Scientific Cluster VSC3.